\documentclass[fleqn,usenatbib]{mnras}
\usepackage[T1]{fontenc}
\usepackage{ae,aecompl}
\usepackage{latexsym,amssymb}
\usepackage{graphicx}
\usepackage{ulem}

\title[Self-regulating jets in Common Envelopes]{\textcolor{black}{Self-regulating jets during the common-envelope phase}}

\author[L\'opez-C\'amara et al.]{
Diego L\'opez-C\'amara$^{1}$\thanks{E-mail: diego@astro.unam.mx},
Fabio De Colle$^{2}$,
Enrique Moreno M\'endez$^{3}$
\\
$^{1}$CONACyT - Instituto de Astronom\'ia, Universidad Nacional Aut{\'o}noma de M{\'e}xico, A. P. 70-264 04510 CDMX, Mexico\\
$^{2}$Instituto de Ciencias Nucleares, Universidad Nacional Aut\'onoma de M\'exico, A. P. 70-543 04510 CDMX Mexico\\
$^{3}$Facultad de Ciencias, Universidad Nacional Aut{\'o}noma de M{\'e}xico, A. P. 70-543 04510 CDMX Mexico}

\date{Accepted XXX. Received YYY; in original form ZZZ}

\pubyear{2018}

\begin{document}
\label{firstpage}
\pagerange{\pageref{firstpage}--\pageref{lastpage}}
\maketitle

\begin{abstract}
Jets launched from a compact object (CO) during a common envelope (CE) may play a key role in the evolution of the system, and may also be an efficient removal channel for its material. In this work we study, through a large set of three-dimensional hydrodynamic simulations, the effects that jets launched from either a \textcolor{black}{black hole (BH)} or a \textcolor{black}{neutron star (NS)} have during a CE phase. \textcolor{black}{The jet is self-consistently powered by a fraction ($\eta$) of the mass accretion rate that reaches the CO.} \textcolor{black}{For low mass accretion efficiencies ($\eta<$0.1\%), the jet is not able to drill through the material accreting onto the CO and forms a bulge around it}. \textcolor{black}{For higher efficiencies, the jet is able to drill through the bulge and in intermediate efficiencies ($\eta\sim$1-5\%) may produce an oscillating mass accretion rate behavior.} We find that the jet \textcolor{black}{may deposit} enough energy to unbind the outer layers of the CE. The \textcolor{black}{self-regulated} jets present variability in size and orientation while their cocoons expand smoothly over the CE. The mass accretion rate initially decreases due to the ram pressure of the cocoon. \textcolor{black}{If the launched jet is not able to drill through the bulge, the accretion is such that it may convert a NS into a BH within a
decade or double the mass of a BH in a few years}. If a jet is present, it \textcolor{black}{deposits} enough energy to unbind the outer layers of the CE and the system \textcolor{black}{evolves} into a grazing envelope, configuration.
\end{abstract}


\section{Introduction}
\label{sec:int}
\textcolor{black}{Many of the transient high-energy phenomena in the Universe are the byproducts of stellar binary interactions in which the common envelope (CE) phase takes place \citep{fryer99, ivanova13}, e.g. SNe Ia (single degenerate: \citealt{whelan73}, double degenerate: \citealt{iben84, webbink84}), ULXs \citep{rappaport2005}, X-ray binaries \citep{tauris06}, short GRBs \citep{berger14}, GWs (double-BH merger: \citealt{abbott16}; double-NS merger: \citealt{abbott2017, holgado18}).} Thus, understanding the dynamics and evolution of the CE phase is a key ingredient in order to determine, both, the rates and the evolution channels in binary systems that produce these high-energy transients.

\textcolor{black}{The CE phase is a short lived stage in the evolution of a binary system in which the envelope of a donor star fills its Roche lobe (RL)} \textcolor{black}{as well as} \textcolor{black}{that of the secondary star} \citep{paczynski76}. Our understanding of the CE phase is not yet complete and even some of the basic questions remain to be fully understood, for instance, \textcolor{black}{the dominating mechanisms responsible for the evolution of the CE phase and its termination} \textcolor{black}{are yet to be determined}. \textcolor{black}{Several approaches have been suggested e.g.:} that a \textcolor{black}{fraction of the initial orbital energy} of the binary system is responsible for \textcolor{black}{its} evolution and termination \citep[the energy formalism of][]{heuvel76}; the $\alpha$-$\lambda$ prescription in which apart from the orbital energy the structure of the envelope is also taken into account \citep{kool90, dewi00}; or the $\gamma$ formalism in which the conservation of angular momentum plays a fundamental role \citep{nelemans00}. \textcolor{black}{Other ingredients which have been considered} \textcolor{black}{to describe the CE phase} are \textcolor{black}{the gravitational potential of the binary and the specific internal energy of the CE \citep{han94}}, tidal heating \citep{iben93}, nuclear energy \citep{ivanova2003}, accretion energy \citep{voss03, soker04}, recombination energy \citep{grichener18}, magnetic field effects \citep{regos95}, and the addition of the enthalpy \citep{ivanova11}. None of the mentioned models have been able to solely account for the driving and ending of the CE phase, nor have they been able to dictate the final configuration or the evolutionary channels of the binary system for certain \citep[see][for  details]{tauris11}.

\textcolor{black}{Given that most massive stars are in binary systems \citep[$\sim$ 70\%,][]{sana12}, there may be binaries in which a compact object (CO) ends up close to a massive star.} As massive stars evolve and enter the giant phase, \textcolor{black}{they may} fill their RL and produce a CE phase with an engulfed CO. Once inside the CE, the CO spirals-in towards the massive star core and accretes material as it moves through the stellar envelope. If the accreted material forms a disk, collimated outflows may be produced. 

Three-dimensional (3D) hydrodynamical (HD) studies of the CE phase have been carried out by previous groups \citep{livio88, terman94, rasio96, sandquist98, sandquist00, demarco03, ricker08, taam10, passy12, ricker12, nandez14, macleod15, ivanova16, kuruwita16, ohlman16a, ohlman16b, staff16, bobrick17, galaviz17, iaconi17a, mmlcdc17, murguia17, shiber17, shiber18, chamandy18}. Due to numerical limitations, it is not yet possible to follow the entire CE evolution phase in a single simulation, but it has been proposed that jets launched within a CE may be an efficient removal channel \citep[e.g.,][]{livio99, armitage00, soker04, papish15}. The energy deposited by the jet may also drive the ejection of the envelope, as described by \citealt{soker16} as the ``jet feedback mechanism'' (JFM). Recently, \citealt{mmlcdc17} presented a 3D HD study of the propagation of a jet launched from a disk located inside the CE, and \citealt{shiber17, shiber18, abubacker18} studied numerically the removal of the CE produced by its interaction with a ``grazing envelope'' (GE, i.e. a configuration in which the jet is located at the edge of the CE and continuously removes the external layers). In these simulations however, the jet is launched with a constant luminosity.

\textcolor{black}{Previous studies have shown that the presence of a jet modifies the final outcome of the CE.} Thus, we follow the dynamics of 3D self-regulated jets, launched from a CO (either a NS or a BH) within a CE, and show how the jets \textcolor{black}{may affect} the outcome of the CE phase. The cocoon, formed as a result of the energy deposited by the jet into the environment, will reduce the accretion rate onto the CO (the ``negative cycle'' of the JFM - see \citealt{soker16} - hereafter NJF). If the $\dot{M}_{\rm{CO}}$ varies with time, so will the ram pressure of the launched jet, and its evolution will be different from a jet that is powered constantly at all times independently of the $\dot{M}_{\rm{CO}}$. \textcolor{black}{For this reason, in this paper} we self-consistently compute the jet power according to the mass accretion rate that reaches the CO \textcolor{black}{($\dot{M}_{\rm{CO}}$)} and assume that the jet is powered by a fraction $\eta$ of the obtained mass accretion rate. 

The paper is organized as follows. In Section~\ref{sec:input}, we describe the characteristics of the models and the numerical simulation parameters. In Section~\ref{sec:results}, we verify the simulation setup, and \textcolor{black}{present the results for an accreting} CO which is either a BH or a NS. In Section~\ref{sec:dis}, we discuss the results. \textcolor{black}{The conclusions are presented in Section~\ref{sec:con}.}

\section{Models and parameters of the simulations.}
\label{sec:input}
The evolution of a self-regulating jet, launched from a CO, and propagating through a CE is followed employing the 3D HD code {\it Mezcal} \citep{decolle12}. \textcolor{black}{The initial setup, unless stated otherwise, is the same used by \citealt{mmlcdc17}.} \textcolor{black}{The mass distribution of the CE corresponds to that of a $20$~M$_\odot$ red giant (RG) star, with a density profile that scales as $\rho_{\rm ce} = \rho_0 (a/R_{\odot})^{-2.7}$, where $a$ is the distance \textcolor{black}{from} the center of mass of the RG \citep{papish15, mmlcdc17}.} \textcolor{black}{The CO, which may be either a 5 M$_{\odot}$ BH, or a 1.4 M$_{\odot}$ NS, is located in the orbital plane of the CE at $a=1.1\times$10$^{13}$~cm} and orbits around \textcolor{black}{the center of mass} with \textcolor{black}{Keplerian} velocity $v_{\rm{k}}$. Viewed from the CO, the CE material (to which hereafter we will refer to as ``the wind'') moves towards the CO with velocity $v_{\rm w}=v_{\rm k}=\left(G M_{\rm ce}(a)/ a\right)^{1/2}$, where  $M_{\rm ce}(a)$ is the mass enclosed up to the orbital distance $a$.

The wind \textcolor{black}{is} present in all the domain at $t=0$~s. At $t\geq0$~s, the wind \textcolor{black}{is} injected from the ZY plane boundary \textcolor{black}{(set at $x=x_b=4\times10^{12}$~cm)}, with the progenitor density and \textcolor{black}{the} \textcolor{black}{Keplerian} velocity at that location. The gravitational effects of the CO were taken into account by adding a gravity term (generated by a point mass at the location of the CO) into the momentum and energy equations. \textcolor{black}{As the material that is accreted by the CO by the end of the integration time is $\sim$two orders of magnitude smaller than that of the CO, the mass of the CO is kept constant at all times.} Since $r_{\rm{in}}$ \textcolor{black}{is} much larger than the Schwarzschild radius for \textcolor{black}{either case of the} COs $\left(r_{\rm{in}} \sim 10^5 r_{\rm{g}}\right)$ and the mass swept by the cocoon \textcolor{black}{is} noticeably lower than that of the CO $\left(M_{\rm{CE,swept}} \sim 10^{-2} M_{\rm{CO}}\right)$, then general relativistic effects of the CO \textcolor{black}{are} negligible and thus not included in our calculations.

A collimated, conical jet is launched from the CO once the $\dot{M}_{\rm{CO}}$ reached a \textcolor{black}{quasi-steady state}. The jet \textcolor{black}{is} launched perpendicular to the orbital and unless stated from the inner boundary $r_{\rm{in}}$=10$^{11}$~cm. \textcolor{black}{The value of $r_{\rm{in}}$ was chosen in order to properly resolve the inner region and a large fraction of the CE for a long integration time.} \textcolor{black}{An extra model used $r_{\rm{in}}$=2$\times$10$^{11}$~cm to show that the results were not sensitive to the value of $r_{\rm{in}}$ a the evolution, global morphology, and $\dot{\rm{M}}_{\rm{CO}}$ were not significantly modified. The expansion velocity of the cocoon and $\dot{\rm{M}}_{\rm{CO}}$ of model BH0.02r$_i$x2 were at most $\sim$20\% larger and $\pm$20\% different (respectively), than the results from model BH0.02long.} The jets velocity and opening angle were $v_{\rm{j}}=c/3$ and $\theta_{\rm{j}}=15^{\circ}$. \textcolor{black}{$\theta_{\rm{j}}$ was chosen to be as narrow as possible within the numerical limitations), and since the dynamics of the jet is dominated by its ram pressure ($\rho_j v_j^2$), different choices of $v_j$ and $\theta_{\rm{j}}$ are equivalent to the usage of different $\eta$ values.}

\textcolor{black}{Unlike \citealt{mmlcdc17}, we assume that the self-regulating jet was powered by a fraction $\eta$ of $\dot{\rm{M}}_{\rm{CO}}$. This is}:
\begin{equation}
\textcolor{black}{\eta \equiv L_{\rm{j}}/({\dot{\rm{M}}}_{\rm{CO}}v_j^2)},
\label{eq:eta}
\end{equation}
\textcolor{black}{where $L_{\rm{j}}$ is the luminosity of the injected jet.} The density of the injected jet \textcolor{black}{is} $\rho_{\rm{j}}= \eta  \dot{M}_{\rm{CO}} / ({\rm{4}} \pi {\rm{r}}^2_{\rm{in}} v_{\rm{j}}$). We assume that, once the material is accreted at the inner boundary, it \textcolor{black}{changes} immediately the power of the self-regulating jet, i.e. we take $t_{{\rm{lag}}}=0.0$. \textcolor{black}{In the appendix A}, we show that setting $t_{{\rm{lag}}}=0.0$ has negligible effects on the general morphology and overall evolution of the self-regulated jet through the CE. The required time for $\dot{\rm{M}}_{\rm{CO}}$ to reach a \textcolor{black}{quasi-steady state} was $t_{\rm{0}}=2\times 10^5$~s. 

\textcolor{black}{The computational domain covered $x_l=z_l=-4\times10^{12}$~cm and $x_r=z_r=4\times10^{12}$~cm in the equatorial plane, while $y_l=$0 and $y_r=$8$\times10^{12}$~cm along the polar axis. We also run models with a smaller computational domain. In these models all the boundaries, except $z_r$, are reduced by a half.} All models \textcolor{black}{have} a reflective boundary set at the equatorial plane, and a constant inflow of material (``the wind'') injected from the ZY boundary plane (set at $x=4\times10^{12}$~cm). All other boundaries \textcolor{black}{have} free outflow conditions. The resolution of the finest level of refinement of all the models was $\Delta = 7.8125 \times 10^{9}$~cm \textcolor{black}{(consistent with the resolution from \citealt{mmlcdc17})} employed where large density, velocity, \textcolor{black}{and} pressure gradients \textcolor{black}{are} present. The resolution of the coarsest level \textcolor{black}{is} $\Delta = 2.5 \times 10^{11}$~cm, employed where no large gradients \textcolor{black}{are} present. The total integration time ($t_{\rm{int}}$) of the models \textcolor{black}{is} between $t_{\rm{int}}=3.00-3.55\times 10^5$~s. To study the long time evolution of the system, we made two simulations with a much larger integration time, $t_{\rm{int}}=8.8\times10^5$~s (long) in a smaller computational domain. \textcolor{black}{In all the simulations, the integration time is not large enough for the jet/cocoon system to reach the wind-injection frontier. Given that the jet/cocoon system reaches the boundaries with outflow conditions supersonically, then the material that crosses any of these boundaries is disconnected from the gas remaining in the domain, thus its loss will not affect the results.}

Each model is labeled according to whether the CO is a NS or a BH, and by the value of $\eta$. If the integration time is long, or if t$_{\rm{lag}}$ \textcolor{black}{is different from zero then this is also included} in the label (\textcolor{black}{as} ``long" or ``lag", respectively). The label, mass of the CO, $\eta$ value, integration time, domain size (``XYZ"), and whether the lag-time is taken into account, for each of the models is shown in Table~\ref{table1}.

\begin{table}
\caption{Initial conditions of the numerical models.}
\begin{center}
\begin{tabular}{cccccc}
  \hline
  Model & $M_{{\rm{co}}}$ & $\eta$ & t$_{\rm{int}}$ & XYZ & t$_{\rm{lag}}$\\
        & ($M_{\odot}$)   &        & (10$^5$s)      &     & \textcolor{black}{(10$^3$s)} \\  
  \hline
  BH0.00 & 5.0 & 0.00 & 3.55 & Big & 0.0 \\
  BH0.001 & 5.0 & 0.001 & 3.53 & Big & 0.0 \\
  BH0.01 & 5.0 & 0.01 & 3.50 & Big & 0.0 \\
  BH0.02 & 5.0 & 0.02 & 3.00 & Big & 0.0 \\
  BH0.05 & 5.0 & 0.05 & 3.35 & Big & 0.0 \\
  BH0.10 & 5.0 & 0.10 & 3.35 & Big & 0.0 \\    
  NS0.00 & 1.4 & 0.00 & 3.55 & Big & 0.0 \\  
  NS0.001 & 1.4 & 0.001 & 3.55 & Big & 0.0 \\
  NS0.01 & 1.4 & 0.01 & 3.52 & Big & 0.0 \\
  NS0.02 & 1.4 & 0.02 & 3.36 & Big & 0.0 \\
  NS0.05 & 1.4 & 0.05 & 3.37 & Big & 0.0 \\
  NS0.10 & 1.4 & 0.10 & 3.37 & Big & 0.0 \\
  NS0.01long & 1.4 & 0.01 & 8.80 & Small & 0.0 \\  
  NS0.05long & 1.4 & 0.05 & 8.80 & Small & 0.0 \\ 
  BH0.02long & 5.0 & 0.02 & 8.80 & Small & 0.0 \\   
  \textcolor{black}{BH0.02r$_i$x2*} & \textcolor{black}{5.0} & \textcolor{black}{0.02} & \textcolor{black}{8.80} & \textcolor{black}{Small} & \textcolor{black}{0.0} \\   
  NS0.05lag & 1.4 & 0.05 & 3.40 & Big & 1.0 \\
  BH0.05lag & 5.0 & 0.05 & 3.42 & Big & 1.0 \\     
  \hline
  \textcolor{black}{*$r_{\rm{in}}$=2$\times$10$^{11}$~cm} \\
  \hline
\end{tabular}
\end{center}
\label{table1}
\end{table}

\section{Self-regulating jets from Compact Objects}
\label{sec:results}
\subsection{{\bf Setup verification}}
\label{sec:verif}
In order to verify that the initial setup was correct and that no numerical artifacts were present, we first confirmed that without a jet the system would reach the quasi-steady wind-accretion solution obtained by Bondi, Hoyle, \& Littleton \citep[BHL,][]{hl39, bh44}. Regardless of the mass of the CO, the $\dot{M}_{\rm CO}$ stabilizes for times larger than \textcolor{black}{$t_{\rm{0}}$}=2$\times10^5$~s \textcolor{black}{(as can be seen in Figure~\ref{fig4} for the cases with $\eta \leq 0.001$).} At \textcolor{black}{t>$t_{\rm{0}}$}, due to the gravitational pull of the CO, the wind material forms a dense elongated bulge (``the BHL bulge'') along the equatorial plane and towards the back of the CO (with respect to the direction of the incoming wind). \textcolor{black}{As \citealt{mmlcdc17} found, the BHL bulge has density values $>10^{-6}$~g~cm$^{-3}$ (with a mean density of order $\sim10^{-5}$~g~cm$^{-3}$), which by t>$t_{\rm{0}}$ completely surrounds the CO.}

The obtained mass accretion rates are smaller than $\dot{M}_{\rm BHL}$ by an $\sim$order of magnitude ($\dot{M}_{\rm CO}\sim 0.1 \dot{M}_{\rm BHL}$). This is due to the density gradient of the RG, and with which, the assumptions of the BHL solution are violated \citep{edgar04}. \textcolor{black}{The $\dot{M}_{\rm CO}$ is also broadly consistent with the rates obtained by previous groups which have studied the CE evolution \citep{ricker08, ricker12, macleod15, macleod17, chamandy18}, as well as with the numerical study of \citealt{beckmann18} were the accretion rate of an adiabatic BHL flow onto a BH are well below the analytic BHL solution.} 

\subsection{Global morphology and evolution of the self-regulated jet within the CE}
\label{sec:global}
\begin{figure*}
   \includegraphics[width=0.9\textwidth]{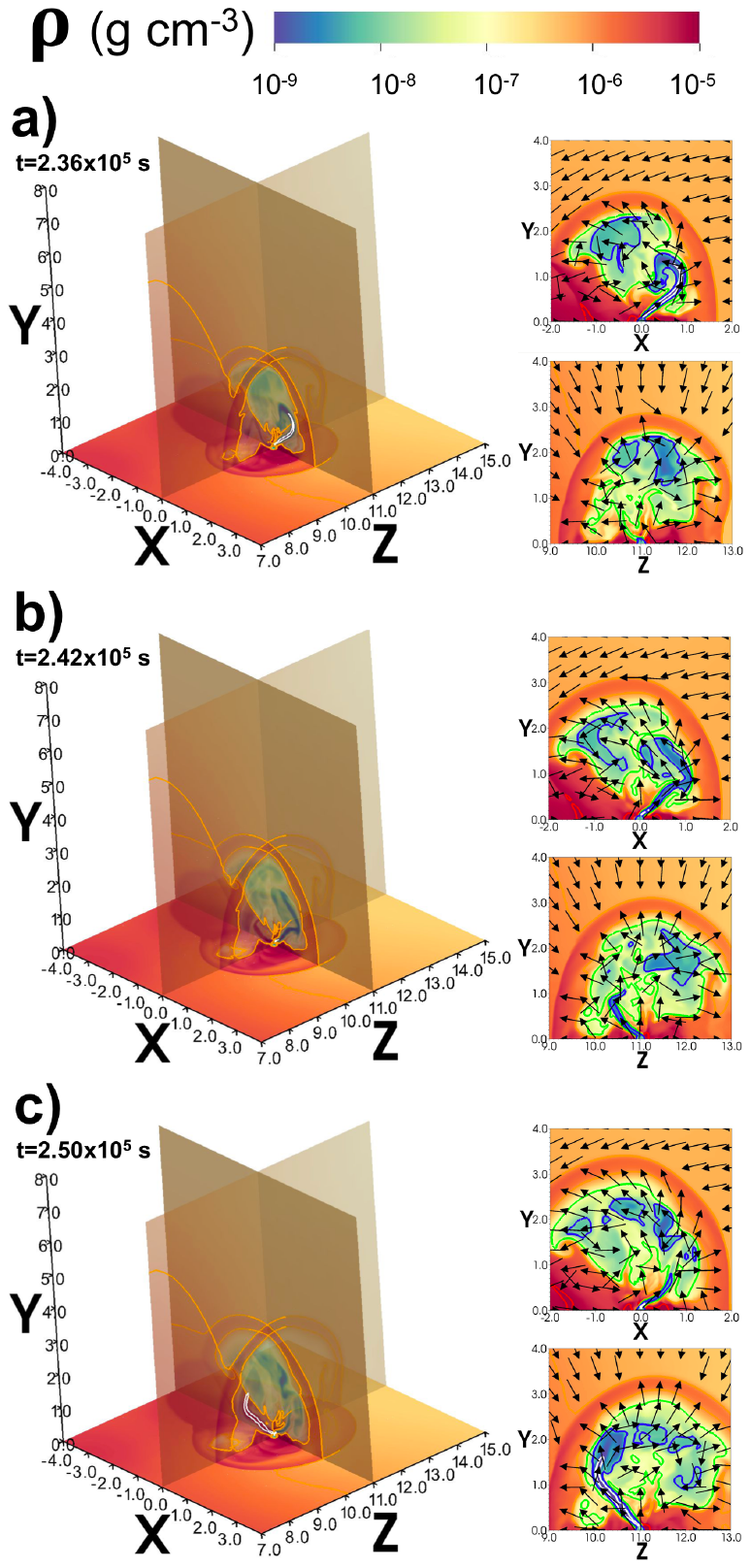}
   \caption{Left panels: XZ, XY, and YZ density map slices showing the 3D time evolution of the CO-self regulated jet-wind evolution ($2.36\times 10^{5}$~s, $2.42 \times 10^{5}$~s, and $2.50 \times 10^{5}$~s for model BH0.05). Right panels: zoom of the density and velocity maps of the XY, and YZ slices. The orange, green, blue, and white isocontour lines correspond to densities $10^{-6}$, $10^{-7}$, $10^{-8}$, and $10^{-9}$~g~cm$^{-3}$, respectively. The axis are in units of $10^{12}$~cm.}
   \label{fig1}
\end{figure*}

The propagation through the stellar envelope of a self-regulated jet launched by a 5~M$_\odot$ BH \textcolor{black}{with $\eta=0.05$ (model BH0.05), is shown in Figure~\ref{fig1} (the figure shows density map slices as well as density map/velocity field slices)}. To describe the morphology and evolution of the jet from model BH0.05, we separated its components into a \textcolor{black}{i)} low-dense region (with fast material, and identifiable by the white isocontour with $\rho = 10^{-9}$~g~cm$^{-3}$), and a \textcolor{black}{ii)} higher-density region (with slower material, blue isocontour with $\rho = 10^{-8}$~g~cm$^{-3}$). At $t=2.36 \times 10^5$~s (panel $a$ of Figure~\ref{fig1}) the cocoon (composed mainly by slowly moving material with $\sim10^{-8}$~g~cm$^{-3}$ and \textcolor{black}{unstable} motion), was already quasi-spherical, \textcolor{black}{it resides} upon the BHL bulge and had been able to expand up to a height $y \sim 2.8 \times 10^{12}$~cm. \textcolor{black}{By this time, the jet (composed by fast-moving material) has been able to drill through the BHL bulge, and moves with an inclination angle of $\sim 45^\circ$ along the XY plane against the wind (consistently with the results of \citealt{mmlcdc17}).} The \textcolor{black}{unstable} motion of the material within the cocoon, as well as the intermittent behaviour of the jet, \textcolor{black}{are clearly visible in Figure~\ref{fig1}}. Twelve and fourteen hours after the launch, the cocoon from model BH0.05 has expanded through the CE maintaining basically the same quasi-spherical structure and reaches $y\sim 3.2 \times 10^{12}$~cm (panel b and c of Figure~\ref{fig1}, respectively). For both times, the slow moving material inside the cocoon still exhibits \textcolor{black}{unstable} motion while the jet \textcolor{black}{varies its size and orientation}. \textcolor{black}{From $t=2.36 \times 10^5$~s up to the total integration time ($t_{\rm{int}}=3.3\times 10^{5}$~s), the cocoon expands smoothly over the CE reaching $y\sim 6 \times 10^{12}$~cm. Meanwhile, the jet remains with a $\sim 45^\circ$ orientation along the ZY, presents variability in its size (varying $\sim \pm$20\% at most), and reaches a maximum length of $y\sim 3 \times 10^{12}$~cm.}

\begin{figure*}
   \includegraphics[width=0.6\textwidth]{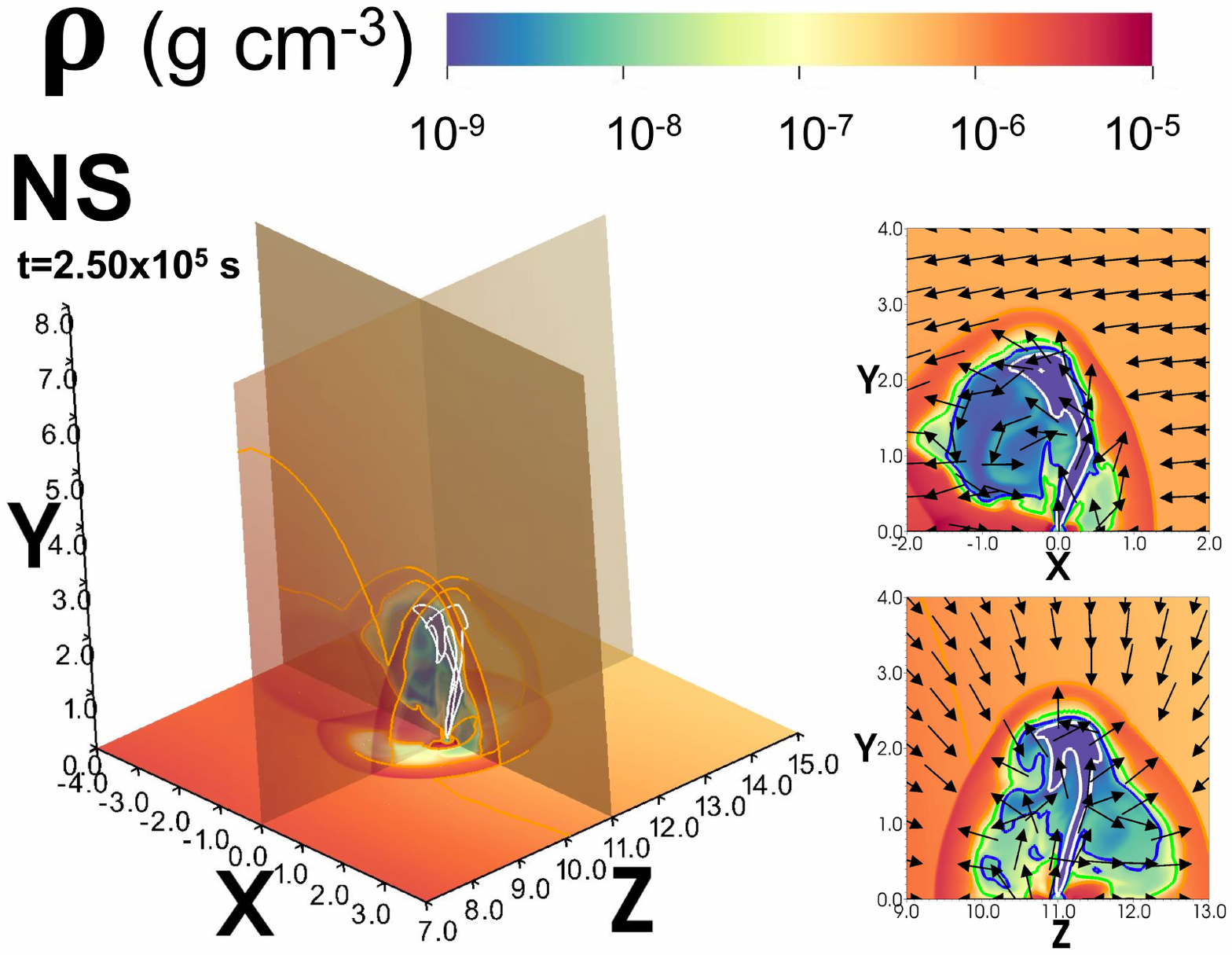}
   \caption{Same as Figure~\ref{fig1} but for model NS0.05.}
   \label{fig2}
\end{figure*}

We also followed the propagation through the stellar envelope of a self-regulated jet launched from a 1.4~M$_\odot$ NS \textcolor{black}{with $\eta=0.05$ (model NS0.05)}. Apart from the fact that the BHL bulge in the NS models is smaller than that from the BH models (as the mass and thus gravitational pull of the NS is smaller to that of the BH), the global morphology and evolution of the NS models is akin to that for the BH models. \textcolor{black}{Figure~\ref{fig2} shows the} density map slices and velocity field slices of model NS0.05. In this case, the cocoon is composed by slow moving material with densities below $10^{-8}$~g~cm$^{-3}$ and \textcolor{black}{unstable} material. By $t=2.50 \times 10^5$~s the cocoon is quasi-spherical, \textcolor{black}{resides} upon the BHL bulge, and had been able to expand up to a height $y \sim 2.8 \times 10^{12}$~cm. Meanwhile, the jet has drilled through its correspondent BHL bulge and is present in both the XY and ZY planes. Due to the reduced gravitational pull of the NS (compared to the BH models), the formed BHL bulge is less dense and smaller than that of the BH models, hence, the forward push exerted by the BHL bulge onto the jet of the NS models is small $\sim 5^\circ$ \textcolor{black}{(compared to the $\sim 45^\circ$ push in the BH models)}. 

\begin{figure*}
 \centering
   \includegraphics[width=0.65\textwidth]{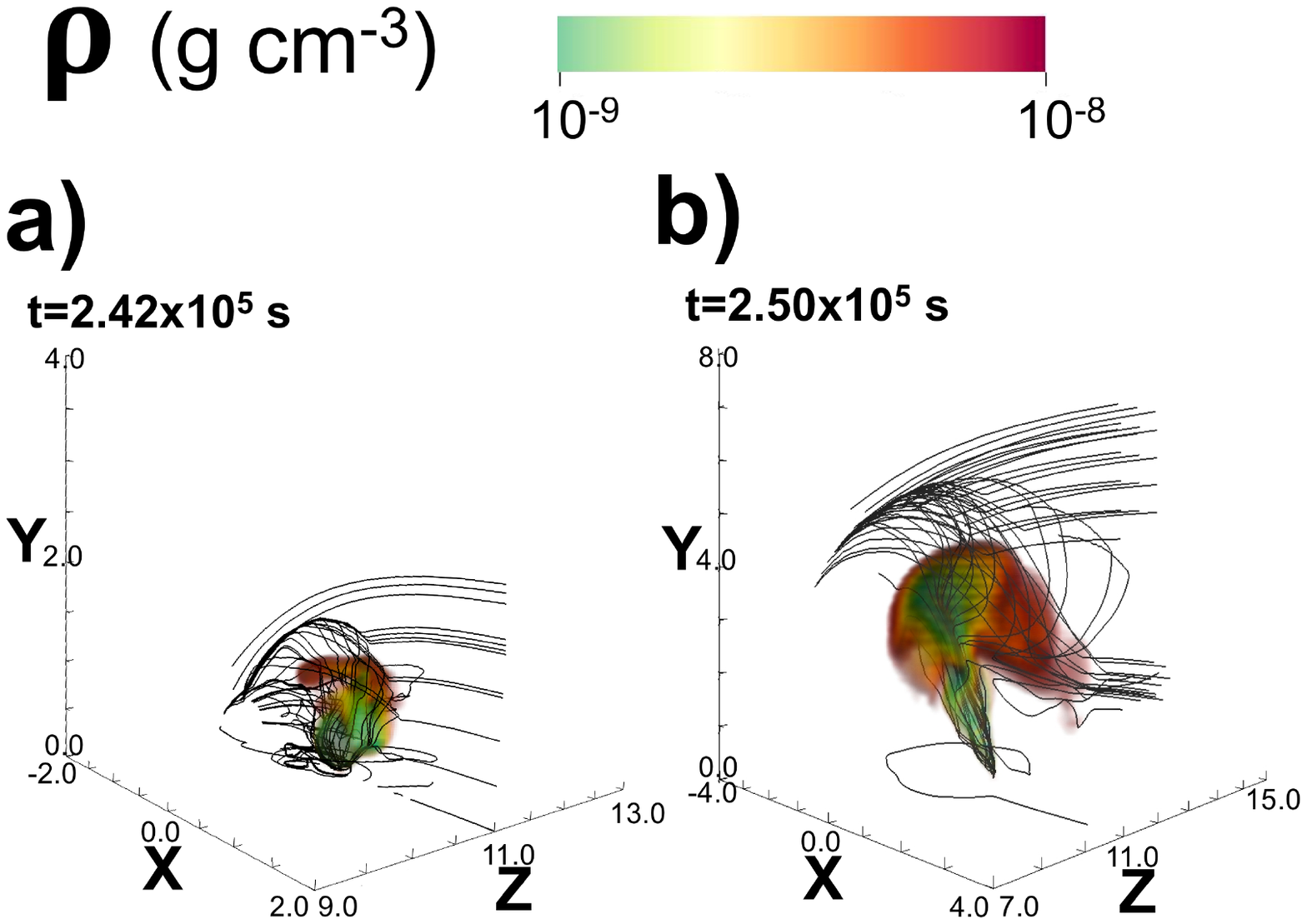}
  \caption{Density volume rendering and velocity stream lines for model BH0.05 at $t=2.42\times 10^{5}$~s (left) and $t=2.50\times 10^{5}$~s (right). The density volume palette covers values from 10$^{-9}$~g~cm$^{-3}$ (green) to 10$^{-8}$~g~cm$^{-3}$ (red). In order to better visualize the stream lines, the  dominion is limited to: $x\in[-2,2]$, $z\in[10,12]$, $y\in[0,2]$. \textcolor{black}{The shaded area represents the XY plane at Z=1.1$\times$10$^{13}$ cm in panel a), and the YZ plane at X=0.0 cm in panel b).} The axis units are the same as in Figure~\ref{fig1}.}
  \label{fig3}
\end{figure*}

\textcolor{black}{In order to understand the size and orientation modification of the jet, in Figure~\ref{fig3} we show the density volume render (showing the less-dense and fastest component of the flow)} and velocity stream lines for model BH0.05. \textcolor{black}{Panel a) corresponds to the time t= 2.42$\times$10$^5$s, and panel b) corresponds to the time t= 2.50$\times$10$^5$s.} Comparing both times, it is clear how \textcolor{black}{the length of the fast component of the flow varies}, and has modified its orientation and has changed its alignment from being mostly in the XY plane to mostly in the ZY plane. \textcolor{black}{This is due to a combination of three different effects: the density gradient along the X axis of the stellar envelope of the RG, the density gradient along the Z axis of the BHL bulge, and the flickering behavior of the self-regulated jet. For the case of the BH, the dominant perturbation is the density gradient of the RG so the jet mainly tilts towards the X axis. For the case of the NS, the dominant perturbation is the density gradient of the BHL bulge, thus the jet tilts against the wind.} The variability of the self-regulated jet is a consequence of the NJF which in turn depends solely on the accretion rate onto the CO and which will be further discussed in Section~\ref{sec:dis}. 

\subsection{Accretion efficiency effects on the jet evolution}
\label{sec:eta}
The value of the efficiency $\eta$, with which a fraction of the accreted mass is redirected into powering the self-regulated jet, has deep implications on the outcome of the CE phase. Depending on the $\eta$ value, the jet will require more or less time to drill through the BHL bulge and the ram power of the jet-cocoon system will be higher or lower. \textcolor{black}{Once the jet breaks out of the BHL bulge, the $\dot{M}_{\rm{CO}}$ diminishes due to the ram pressure of the jet-cocoon (i.e. due to the NJF).} Independently of the CO, if the self-regulated jet is able to drill through the BHL bulge, it will then evolve through the stellar envelope and produce a cocoon that surrounds the jet and expands smoothly inside the CE. Small $\eta$ values produce self-regulating jets with low ram pressure \textcolor{black}{which} may not be able to break out of the BHL bulge or may be quenched by the accreting material (see Section~\ref{sec:dis} for further discussion). Independently of the CO, the models with $\eta<0.001$ were never able to drill through the BHL bulge, the model with $\eta=0.01$ required $\sim10^5$~s to drill through the BHL bulge, while the models with $\eta \gtrsim 0.02$ were able to drill through the BHL. Models with very high $\eta$ values on the other hand, present basically the same jet-cocoon morphology as those from their respective models with $\eta=0.05$ (see Section~\ref{sec:global}). Larger $\eta$ efficiency produce jets with larger ram pressures, thus, the correspondent self-regulated jet and cocoon will advance faster through the CE (e.g. model BH0.10 reached $y\sim8\times10^{12}$~cm by the end of its integration time). 

\begin{figure}
 \centering
   \includegraphics[width=0.48\textwidth]{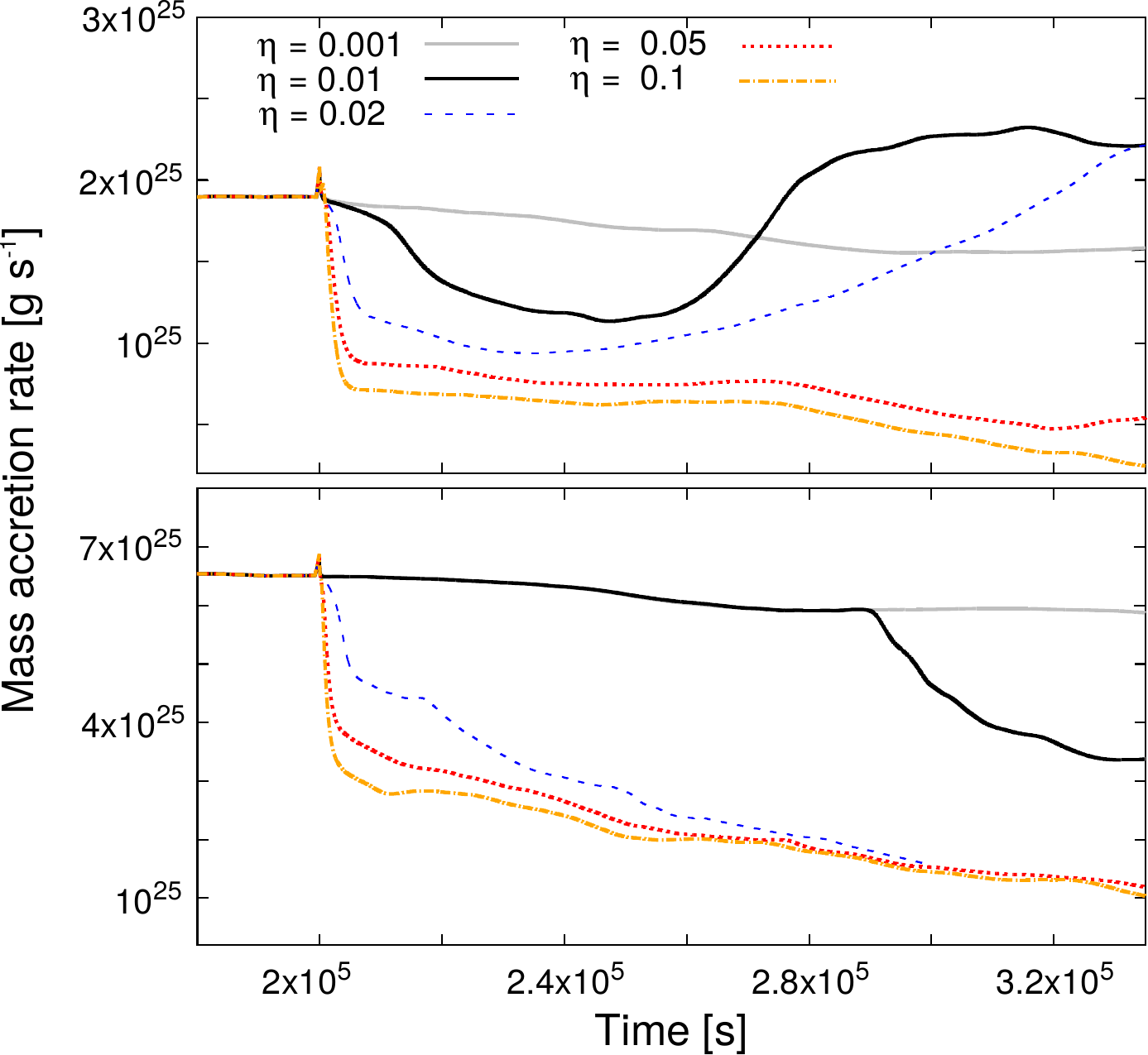}
  \caption{Mass accretion rate onto the CO for different values of $\eta$ (0.001 grey solid line, 0.01 black solid line, 0.02 blue dashed, 0.05 red dotted, 0.10 green dot-dashed) and mass of the CO ($M=1.4$~$M_\odot$ in the upper panel and $M=5$~$M_\odot$ in the bottom panel - see also Table 1) as a function of time.}
  \label{fig4}
\end{figure}

Figure~\ref{fig4} shows $\dot{M}_{\rm{CO}}$ as a function of time for all the models \textcolor{black}{(upper panel for the NS models, and bottom panel for the BH models)}. \textcolor{black}{For all cases,} before the jet is launched, we let the system reach a quasi-steady state by $t=2 \times 10^5$~s in which $\dot{M}_{\rm{CO}} \approx 2 \times 10^{25}$~g~$\textcolor{black}{s^{-1}}$ for the NS models and $\dot{M}_{\rm{CO}} \approx 6.5 \times 10^{25}$~g~$\textcolor{black}{s^{-1}}$ for the BH models. \textcolor{black}{The $\dot{M}_{\rm{CO}}$ of model BH0.001 (black line of the bottom panel of Figure~\ref{fig4}) remains constant up to 2.9$\times 10^{5}$~s given the low power of the jet.} The BH models with successful jets diminish the $\dot{M}_{\rm{CO}}$ down to $\sim 3\times 10^{25}$~g~$\textcolor{black}{s^{-1}}$ in a few hours, and then decrease down to $\sim 20$\% of the accretion rate at $t_{\rm{0}}$ by the end of the integration time. The NS models with successful jets also present accretion rates that drop down rapidly to 25-50\% of the accretion rate at $t_{\rm{0}}$. 

The NS models with intermediate accretion efficiencies ($\eta = 0.01-0.02$) present a very interesting behavior. \textcolor{black}{For these, once the self-regulated jets break out of the BHL bulge the accretion rates diminish accordingly to the rest of the high-$\eta$ value models but may then increase to mass accretion rate values that may even be larger than that at $t_{\rm{0}}$.} \textcolor{black}{For example, model NS0.01 has a mass accretion rate at $t_{\rm{0}}$ close to $\sim$1.8$\times$10$^{25}$~g~$\textcolor{black}{s^{-1}}$ which then varies between $\sim$1.1$\times$10$^{25}$~g~$\textcolor{black}{s^{-1}}$ and $\sim$2.3$\times 10^{25}$~g~$\textcolor{black}{s^{-1}}$.} Model NS0.02 meanwhile, after reaching a minimum mass accretion value \textcolor{black}{($\sim 10^{25}$~g~$\textcolor{black}{s^{-1}}$)}, increases $\sim$monotonically up to a mass accretion rate (\textcolor{black}{$\sim$2.2$\times$10$^{25}$~g~s$^{-1}$} which is approximately 20\% larger with respect to the mass accretion rate at $t_{\rm{0}}$. 

\begin{figure}
 \centering
   \includegraphics[width=0.48\textwidth]{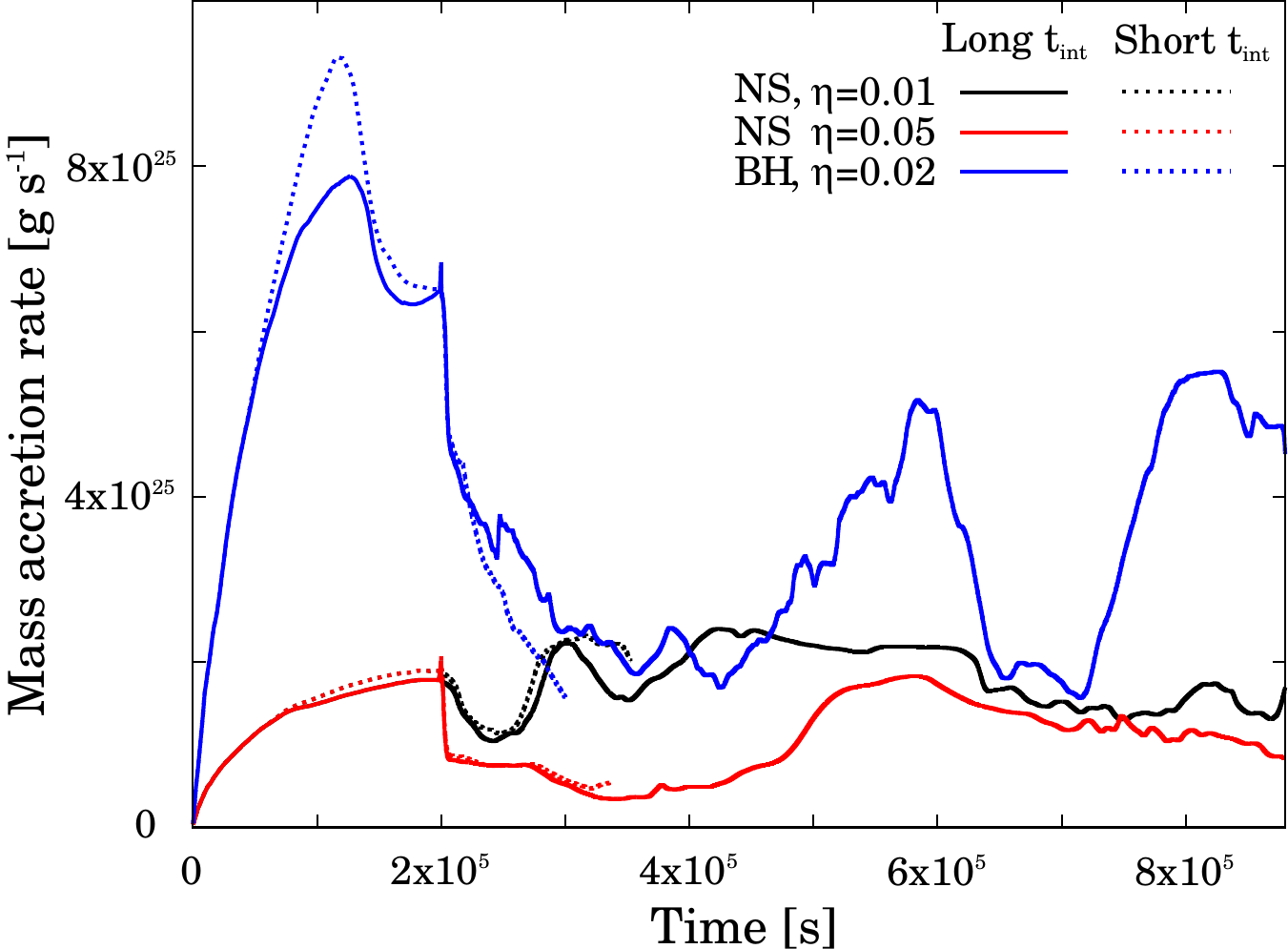}
  \caption{Same as Figure~\ref{fig4}, but for models with long integration times (solid lines) for different models NS0.01long (black), NS0.05long (red), and BH0.02long (blue). The respective models with short integration times are also included (dotted lines).}
  \label{fig5}
\end{figure}

In order to understand the long time evolution of the accretion rate, we ran an extra set of simulations with a much larger integration time (NS0.01long, NS0.05long, and BH0.02long, see Table 1 for further details). Due to the domain reduction in the small domain models \textcolor{black}{(see Section~\ref{sec:input} for further details)}, the accretion rate in these models is somewhat different to that from the large domain models since less material from the wind injection boundary is followed. The $\dot{M}_{\rm{CO}}$ from the long integration time models are shown in Figure~\ref{fig5}. All three models present variable behavior. \textcolor{black}{Model NS0.01long varies between $\sim$10$^{25}$~g~$\textcolor{black}{s^{-1}}$ and $\sim$2.2$\times 10^{25}$~g~$\textcolor{black}{s^{-1}}$. NS0.05long varies between $\sim$5.0$\times 10^{24}$~g~$\textcolor{black}{s^{-1}}$ and $\sim$1.8$\times 10^{25}$~g~$\textcolor{black}{s^{-1}}$. Model BH0.02long, varies between $\sim$2.0$\times 10^{25}$~g~$\textcolor{black}{s^{-1}}$ and $\sim$8.0$\times 10^{25}$~g~$\textcolor{black}{s^{-1}}$.}

\section{Discussion}
\label{sec:dis}
In this section, we first discuss the evolutionary paths leading to the binary systems considered in this paper. Then, guided by the results of the numerical simulations, we discuss the role that the NJF and self-regulating jets might play in the evolution and \textcolor{black}{termination} of the CE phase. Finally, we briefly examine the implications of our results for the formation of double CO binaries.

\subsection{Evolutionary channels}
\label{sec:evol}
\textcolor{black}{In our models, we considered a binary system formed by a massive RG and a CO (either a BH or a NS, each with a few M$_\odot$).} The system formed by a massive star and a NS is, most likely, the result of the evolution channel of a binary system formed by two massive stars (the primary with a mass between $M=10-20M_\odot$, and the secondary between $M=8-16 M_\odot$), and an orbital separation of a few AU. The interaction between the two stars is weak until the primary star forms a large He core (in order to guarantee the formation of a NS). During its evolution, the primary star fills its RL and transfers mass through the Lagrangian point ``L1" onto the secondary star. Once the primary explodes as a supernova (SN), a NS is produced, and the secondary may have accreted a large fraction of the envelope of the primary star ending up as a RG with $\sim 20 M_\odot$. Meanwhile, the system formed by a massive star and a BH is most likely produced by the interaction of a $M\sim40 M_\odot$ primary, a secondary with $M \sim 20 M_\odot$, and an orbital separation of $\sim 10$~AU. In this case, after the primary star explodes as a SN, it produces a $5M_\odot$~BH and a RG with $\sim 20 M_\odot$. In both cases, two main requirements must be met in order to avoid unbinding of the system. First, the SN kick must be small and it should preferably be directed against the orbital velocity of the binary system \citep{podsiadlowski05}. Second, the mass lost from the system during the SN explosion must be smaller than the mass remaining in the system \citep{blaauw61, boersma61}. Also for both cases, due to the mass transfer from the primary to the secondary star (during the Roche lobe overflow of the primary star), the orbital separation of the binary system will be reduced to $\sim 10^{13}$~cm. Thus, when the secondary star evolves into its RG stage, there will be a second phase of unstable mass transfer of the material of the secondary star through L1 onto the CO (we will refer to the secondary star as the ``donor"), and the CO will be engulfed within the stellar envelope. The CO will then accrete material of the donor, may form an accretion disk, and in turn could power a jet, which kinetic luminosity will change with time depending on the $\dot{M}_{\rm{CO}}$.

\begin{figure}
 \centering
   \includegraphics[width=0.48\textwidth]{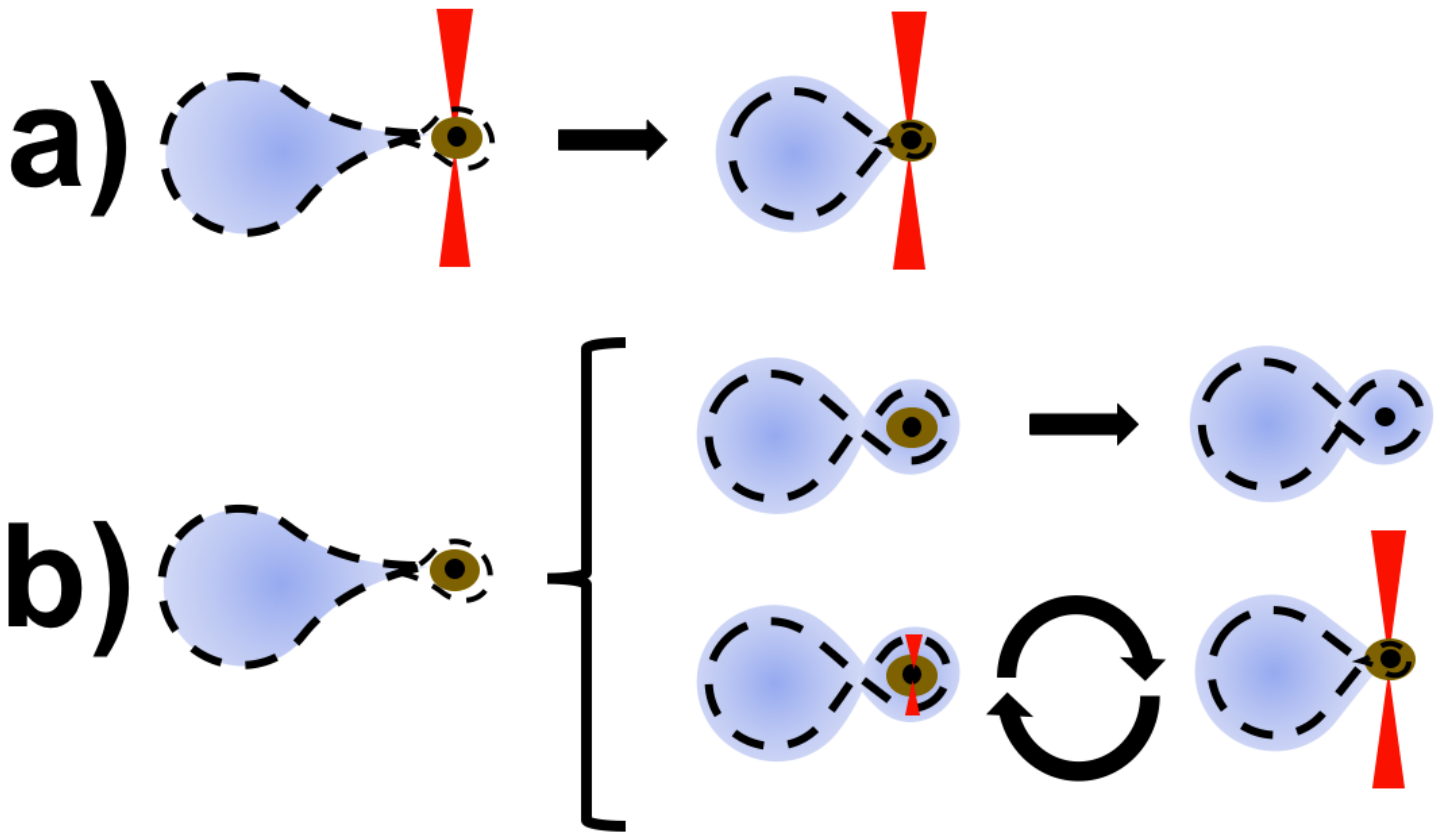}
  \caption{Schematic representation of the possible outcomes resulting from the interaction of the common envelope with a self-regulating jet (see text for discussion).}
  \label{fig6}
\end{figure}

\subsection{Self-regulating jets in Common Envelopes}
\label{sec:jetCE}
The ejection of a jet from an accretion disk around a CO is an open problem \citep[for further details see][]{bete05}. Some of the most accepted mechanisms are: i) the extraction of energy from a rotating black hole and launching of the jet by the presence of a large-scale magnetic field in the ergosphere \citep{bz77}, ii) extraction of energy from a rotating accretion disk with a frozen magnetic field \citep{bp82}, iii) the launch of a jet due to a star-disk system in interaction \citep{shu94}, and iv) the powering from a fireball produced by the neutrino annihilation from a \textcolor{black}{fraction of the neutrinos} produced in an accretion disks \citep{pwf99}. In all the models a jet may or may not be launched due to a variety of reasons (e.g., low magnetic field, synchronous rotation). 

The evolution of a binary system \textcolor{black}{in which a jet is launched from a CO,} depends dramatically on whether the CO launches a jet before it is engulfed by the CE. The two main evolutionary channels are shown in Figure \ref{fig6}, and are next described in detail (guided by the results of the numerical simulations). First, we consider the case in which the accretion disk/CO system ejects a jet before it is engulfed by the CE. Once the edge of the expanding envelope of the donor arrives at the location of the accretion disk/CO system, the jet may have enough energy to remove the external part of the envelope closest to it (see \citealt{shiber18} for further details). While the mass transfer from the donor to the CO continues, so will the removal of the outermost layers of the donor envelope as the CO \textcolor{black}{orbits} closer to the stellar core of the donor star (\textcolor{black}{case} a from Figure \ref{fig6}). This process will continue until the donor stops filling its own RL, or the jet is switched off. \textcolor{black}{Second}, when the CO does not have an associated jet (or if the jet is not powerful enough) at the beginning of the CE phase it may become fully engulfed by the CE before being able to launch a jet \textcolor{black}{(\textcolor{black}{case} b from Figure \ref{fig6})}. Once the CO is within the CE, it will accrete material at a fraction of the BHL accretion rate \citep{mmlcdc17}:
\begin{equation}
\dot{M}_{\rm BHL} = 10^{26} \left(\frac{\dot{M}_{\rm{CO}}}{1.4 M_\odot}\right)^2 \left(\frac{M(a)}{20 M_\odot}\right)^{-3/2} \left(\frac{a}{10^{13}{\rm cm}}\right)^{-1.2} {\rm g\; s^{-1}}\;,
\label{eq:BHL}
\end{equation}
where $\dot{M}_{\rm{CO}}$ is the mass of the CO, $M(a)$ is the stellar mass up to the radius $a$ \textcolor{black}{of the RG stellar profile taken into account \citep{papish15}}. Consistently with previous studies \citep{macleod15, lora15, mmlcdc17}, our simulations show that the $\dot{M}_{\rm{CO}}$ is $\sim 20$\% of the correspondent BHL mass accretion rate shown in Equation~\ref{eq:BHL}. The presence of a jet in the CE phase changes drastically the $\dot{M}_{\rm{CO}}$. \textcolor{black}{Given that the jet is self-regulated by the mass accretion rate, a drop (or increment) in $\dot{M}_{\rm{CO}}$ will reduce (or increase) L$_{\rm{jet}}$ and the corresponding ram pressure and density.}

Numerical simulations from Section~\ref{sec:results} show that, independently of the CO (NS or BH), if the accretion to ejection efficiency with which the self-regulates jet is powered is $\eta\lesssim 0.001$, the jet does not manage to drill through the dense BHL bulge which surrounds the CO, with which the CO will have an accretion rate similar to the case in which no jet is launched. If the efficiency $\eta\gtrsim 0.01$ (typically, astrophysical jets have an efficiency $\sim 10\%$, e.g., \citealt{livio99}), then the ram pressure and energy deposition of the jet in the CE will produce a NJF which in turn leads to a drop in the $\dot{M}_{\rm{CO}}$. \textcolor{black}{We assume that, except for the material ejected in the jet, all the gas that enters the spherical surface at r$_{in}$ is accreted by the CO. We also suppose that all of this material is accreted $\sim$instantaneously by the CO since the temporal resolution in our simulations ($\Delta t \sim$10$^{2-3}$s), is of order of the lagtime (i.e. the time required by the material that crosses r$_{in}$ to reach the surface of the CO, see the appendix A for further details).}

The NJF and drop in the $\dot{M}_{\rm{CO}}$ can be qualitatively understood as follows. \textcolor{black}{Since the density profile of the RG follows that from \citealt{papish15}, then by assuming that the material of the CE accretes onto the CO at approximately the free-fall velocity $v_{ff}\sim \sqrt{GM_{\rm CO}/R}$ (where R is the distance to the CO, see \citealt{edgar04} for further discussion), the ram pressure of the free-falling material is:}
\begin{equation}
P_{ff} = \rho_{sh} v_{ff}^2 \textcolor{black}{\ \propto} \left(\frac{a}{R_\odot}\right)^{-2.7} \frac{GM_{\rm CO}}{R}
\textcolor{black}{\ \propto} \ a^{-2.7} M_{\rm CO}/R,
\end{equation}
where $\rho_{sh}$ is the density of the shocked envelope material.

As the jet decelerates by its interaction with the environment, part of its kinetic energy is dissipated into thermal energy of the cocoon. Thus, the cocoon pressure ($P_c$, which opposes the accretion of the wind material onto the CO), can be estimated as a fraction of the pressure of the jet. This is:
\begin{equation}
P_c\sim \rho_j v_j^2=\dot{M}_j v_j/(4\pi R^2)\textcolor{black}{\ \propto \ }  \eta M_{\rm CO}^2 M(a)^{-3/2} a^{-1.2}/R^2 \;,
\end{equation}
where $\rho_j$, $v_j$ and $\dot{M}_j$ are the density, velocity and mass ejection rate of the jet, and where we have employed equation~\ref{eq:BHL}.

\textcolor{black}{If the pressure of the cocoon is larger than the free-fall pressure, then the NJF will come into play and the $\dot{M}_{\rm{CO}}$ will stop. This happens when the mass of the CO is greater than a critical value ($M_{\rm CO,crit}$), which is obtained by the condition $P_c > P_{ff}$, this is, when: $M_{\rm CO}> M_{\rm CO,crit}\propto R~a^{-1.5} M(a)^{3/2} \eta^{-1}$.} In the case of accretion onto a CO, the role of the NJF and the drop in the $\dot{M}_{\rm{CO}}$ will depend on the ram pressure of the jet. If $\eta$ is large, so is the ram pressure of the jet, and so is the deposited energy and ram pressure of the cocoon. Thus, the cocoon will stop a vast amount of the shocked material which otherwise would be accreting towards the CO. As the ram pressure of the cocoon scales as $M_{\rm CO}^2$ while the free fall pressure scales as $M_{\rm CO}$, the accretion rate drops for CO with larger mass. 

In the case of a constantly powered jet the $\dot{M}_{\rm{CO}}$ is suppressed up to an order of magnitude in about $\sim$ a day \citep[see][]{mmlcdc17}. On the other hand, if the jet is self-regulated, the drop in the accretion rate is more moderate. This is visible in both Figures~\ref{fig4} and \ref{fig5} where the accretion rate drops at most by a factor of $\sim$ 4 and 6 for the NS and BH case (respectively) once the jet breaks out of the BHL bulge. Initially, the $\dot{M}_{\rm{CO}}$ drops due to the cocoon expansion. As the free-fall pressure is $\propto 1/R$ and the cocoon pressure is $\propto 1/R^2$, then at a certain distance from the CO the free-fall pressure dominates, reviving the $\dot{M}_{\rm{CO}}$ \textcolor{black}{(as can} be seen in Figures~\ref{fig4} and \ref{fig5}). This process might repeat itself producing variability up to a \textcolor{black}{factor of a few}, until the mass supply runs out. It may even leave a bare NS at the end of the CE phase \citep{papish15}, or the NS may be converted into a BH given that the accretion rates are some eight orders of magnitude above the Eddington limit ($\sim$10$^{-1}$M$_{\odot}$yr$^{-1}$, see \textcolor{black}{Section~\ref{sec:ligo} for further discussion}).

Two timescales are key to understand the outcome of the jet-CE interaction: the time the jet takes to break out from the envelope ($t_{\rm{bo,e}}$) and the time needed to unbind the envelope ($t_{\rm{u,e}}$). Once the jet is formed, it moves through the envelope at a reduced velocity (as it is slowed down by the interaction with the environment), of order of $v_{\rm{sh}}\sim 10^{8}$~cm/s according to our simulations. If the CE has a stellar surface with radius $a=10^{13}$~cm, then after about $t_{\rm{bo,e}} \sim$10$^{5}$~s ($\sim$a day) the jet will break out of the  CE. This timescale can be compared with the time required to unbind the outer part of the envelope. Taking a stellar mass given by $M(a)=M_{core}+ M_{env}(a/R_{CE})^{0.3}$ (where $M_{core}=3.5 M_\odot$,$M_{env}=16.5 M_\odot$ and $R_{CE}= 535 R_\odot$ is the CE radius; see \citealt{papish15}),
the binding energy is:
\begin{equation}
 E_{\rm bind} = \int_a^{R_{CE}} \frac{G M(r) dM}{r} \approx 10^{49} {\rm erg}\;,
\end{equation}
at a=1.1$\times$10$^{13}$ cm, while the jet energy is computed as:
\begin{equation}
  E_{\rm jet} = L_{\rm jet} t = \eta \dot{M}_a v_j^2 t = (2-4) \times 10^{44} t\; {\rm erg}\;,
\end{equation}
where we have computed $\dot{M}_a$ directly from the mass accretion rate at the inner boundary from our simulations, $\eta$ is assumed to be ${\rm{10}}^{-1}$, and $v_j=c/3$. Thus, we find that $t_{\rm{u,e}} \sim E_{\rm bind}/ L_{\rm jet} \simeq 2-5 \times 10^4$~s. \textcolor{black}{Considering the dependence of M(a) and $\dot{M_{\rm a}}$ with the radius, we get E$_{\rm{jet}} \propto a^{-1.65}$, and E$_{\rm{bind}} \propto a^{-0.5}$. With these scalings the jet is able to unbind the envelope within hours when a$\sim$10$^{11}$cm, and days if the CO is located at a$>$10$^{11}$cm. That is, within a few minutes to several days (depending on the value of $\eta$ and a), the self-regulated jet deposits enough energy into the envelope to terminate the CE phase.}

Numerical simulations show that in the case of a constantly powered jet the \textcolor{black}{NJF acts at all times and the $\dot{M}_{\rm{CO}}$} drops dramatically. \textcolor{black}{From our simulations of a self-regulated jet, we find that even though the $\dot{M}_{\rm{CO}}$ is intermittent, it is always between 10$^{25}$g~s$^{-1}$ to 10$^{26}$g~s$^{-1}$.} Thus, the energy deposited by the jet during $\sim$ hours to days is sufficient to unbind the outer layers of the CE. \textcolor{black}{We find that a NS located at $a\sim10^{13}$~cm could accrete 10$^{-4}$~M$_\odot$ while the CE lasts (not affecting the orbital separation nor the $\dot{M}_{\rm{CO}}$). The 5 M$_\odot$ BH on the other hand, would accrete 10$^{-3}$~M$_\odot$ while the CE phase lasts. If this phase lasted $\sim$10~yr, with similar $\dot{M}_{\rm{CO}}$ and orbital separation, the NS would collapse into a BH.}

\subsection{{\bf Implications for the formation of double compact object binaries}}
\label{sec:ligo}
\textcolor{black}{The Eddington limit, relevant for spherically symmetric accretion onto a CO in a homogeneous medium \citep[i.e. Bondi accretion]{bondi52}, does not apply in our study since we have BHL accretion with a density gradient onto a thick extended disk. Still, as it is used as a reference for accretion rates, we compare our results to it. Consistently with the values found in \citealt{macleod15, mmlcdc17} the obtained $\dot{M}_{\rm{CO}}$ in our models ($\sim0.3 M_\odot$ per year for a NS) are seven to eight orders of magnitude above the Eddington limit for photons and twelve orders of magnitude below the Eddington limit for neutrinos. This accretion rate is three to four orders of magnitude above the limit where neutrino cooling becomes important \citep{FryerColgate99, moreno11}. This scenario is known as hypercritical accretion in the literature \citep[see, e.g.,][]{brown94, chevalier95, moreno08, moreno11}.}

\textcolor{black}{If no jets are present (or if their power is considerably low and does not manage to drill through the BHL bulge) in the NS-RG system (within the CE), the NS may be converted to a BH, preventing the formation of double-NS (DNS) binaries.} Since we know of at least eleven confirmed DNS systems \citep{heuvel17, tauris17}, and the most promising progenitor of GW/SGRB170817 is a DNS merger \citep{abbott2017, coulter17}, then the accretion scenario in which the NS collapses to a BH may be rare. However, we cannot rule out that the GE \citep[][]{shiber17,shiber18} may produce a more stable mass-transferring system in which the NS does collapse into a BH. If jets are present, then a double-NS (DNS) binary or a NS-BH binary will be formed. \textcolor{black}{The GWs from CE systems consisting of a NS and a massive RG will be detectable by aLIGO for distances up $\sim$50kpc \citep{holgado18}.} If jets are present before the CO is engulfed by the CE, then a GE will appear. In this case, whether the NS collapses or not into a BH depends on the $\dot{M}_{\rm{CO}}$ during the GE evolution. If the jets are launched once the CO is immersed within the CE, then the accreting NS may be converted into a BH before the jets are launched. As the launched jets remove a large fraction of the CE, then the GE case will ensue. Future detections of GWs/GRBs and observed rates of these systems will be very helpful to fully understand the CE phase and its outcomes.

\section{Conclusion}
\label{sec:con}
The simulations presented in this paper illustrate how self-regulated jets play an important role during the CE phase, when a CO is immersed in the stellar envelope of a RG star. Within a few hours to several days after a jet is ejected from the accretion disk/CO system (depending on the efficiency with which the jets are powered), the jet deposits enough energy into the envelope to unbind the outer layers of the CE.

The global morphology and evolution of the jets \textcolor{black}{through the CE} is very similar \textcolor{black}{independently on whether the CO} is a NS or a BH (except for the size and density of the BHL bulge). Depending on the efficiency with which a fraction of the accreted mass \textcolor{black}{powers} the self-regulated jets, \textcolor{black}{the evolution of the CO-RG system within the CE will follow one of these two scenarios}. For low efficiencies \textcolor{black}{($\eta < 0.001$)}, the jet will not manage to drill through the dense BHL bulge. On the other hand, if the efficiency is high enough \textcolor{black}{($\eta < 0.01$)}, the self-regulated jet \textcolor{black}{will be able to drill through the bulge, and then evolve through the CE (with a variable behaviour in both its size and orientation), while its correspondent cocoon} expands smoothly over the CE. 

\textcolor{black}{If the self-regulated jet is able to drill through the BHL bulge}, the mass accretion rate \textcolor{black}{which reaches the} CO will drop due to the ram pressure of the cocoon which competes against \textcolor{black}{that} of the accreting material. Hence, the powering of the jets will decrease until the pressure of the accreting material becomes larger than that of the cocoon, and thus the mass accretion rate onto the CO and powering of the jet will increase once more. Unlike the case where the jets are powered constantly over time (in which the mass accretion rate drops monotonically with time), the self-regulated jets present an oscillating behavior (where the mass accretion rate may increase or decrease by a factor of a few). This process, will repeat itself until the mass supply runs out. Depending on the mass accretion rate onto the CO, if the CO within the CE is a NS it may be converted into a BH in \textcolor{black}{about a} decade (if the CO is a 5 M$_\odot$ BH it will double its mass in a \textcolor{black}{few} years). If jets are launched once the CO is engulfed by the CE, \textcolor{black}{they deposit} enough energy to unbind the outer layers of the \textcolor{black}{CE. Once the outer layers of the CE are blown away, or if the CO presents a jet before it is engulfed by the CE}, a \textcolor{black}{GE configuration may appear}. Future detections of GWs/GRBs and observed rates of these systems will be very helpful to fully understand the CE phase and its outcomes.

\section*{Acknowledgements}
We thank the reviewer for providing helpful comments that improved this paper. D.L.C is supported by C\'atedras CONACyT at the Instituto de Astronom\'ia (UNAM). We acknowledge the support from the Miztli-UNAM supercomputer (projects LANCAD-UNAM-DGTIC-321 and LANCAD-UNAM-DGTIC-281 for the assigned computational time in which the simulations were performed. F.D.C. thanks the UNAM-PAPIIT grants IA103315 and IN117917.


\appendix
\section{{\bf Lag time from the inner boundary to the CO}}
In the numerical simulations considered in this paper, we have assumed that the accreting material that reaches the inner boundary ($r_{\rm{in}}$) immediately alters the jet powering. This is not strictly true since the jet is launched very close to the CO radius ($r_{\rm{CO}}$) which is many orders of magnitude smaller than $r_{\rm{in}}$. The material which crosses $r_{\rm{in}}$ still requires some extra time interval, t$_{\rm{lag}}$, to reach $r_{\rm{CO}}$ and, thus, affect the powering of the jet. In the following, we show that including a $\rm{t}_{\rm{lag}} > 0$ does not affect the outcome of the simulations.

Among all possible trajectories that material crossing $r_{\rm{in}}$ may follow on its way to $r_{\rm{CO}}$, three are particular relevant: i) accretion through the accretion disk \textcolor{black}{only}, ii) a free-falling trajectory that directly falls onto the CO, or iii) a combination of the two in which the material free-falls onto the accretion disk and then spirals-in towards the CO. 

If the material is moving in the equatorial plane and an accretion disk has already formed and has a radius r$_{\rm{d}}\gtrsim$r$_{\rm{in}}$, then the material will spiral inwards towards the CO through the accretion disk in a viscous timescale (t$_{\rm{visc}}$). The viscous timescale for the material that is in an accretion disk spiraling inwards towards the CO, is:
\begin{equation}
\rm{t}_{\rm{visc}} = \frac{{\rm{r_d}}}{{\rm{v_R}}},
\label{eq1}
\end{equation}
where r$_{\rm{d}}$ is the radius of the disk, and $\rm{v_R}$ the radial velocity of the material. Taking into account the kinematic viscosity of the accretion disk, and assuming that the disk is thin (H $\ll$ r$_{\rm{d}}$, where H is the disk scale height), then the $\rm{v_R}$ can be obtained as:
\begin{equation}
{\rm{v_R}} \approx \frac{\nu}{\rm{r_d}}.
\label{eq2}
\end{equation} 

Considering the $\alpha$-prescription of \citealt{shakura73} for a thin disk, the kinematic viscosity is:
\begin{equation}
\nu = \alpha c_s {\rm{H}},
\label{eq3}
\end{equation}
where $\alpha$ is a value between 0 and 1, and $c_s$ the sound speed in the disk. 

If the disk is in hydrostatic equilibrium, and is dominated by the thermal pressure, then the H and $c_s$ of an $\alpha$-prescription accretion disk is (see \citealt{fkr02} for more details):
\begin{equation}
{\rm{H}} = 1.16 \times 10^7~\alpha^{-1/10}~{\dot{\rm{M}}}_{\rm{in}}^{3/20}~{{\rm{M_{CO}}}}^{-3/8}~{{\rm{r_d}}}^{9/8},
\label{eq4}
\end{equation}

\begin{equation}
c_s  = 2.11 \times 10^3~\alpha^{-1/10}~{\dot{\rm{M}}}_{\rm{in}}^{3/20}~{{\rm{M_{CO}}}}^{1/8}~{{\rm{r_d}}}^{-3/8},
\label{eq5}
\end{equation}
where M$_{\rm{CO}}$ is the mass of the CO, and ${\dot{\rm{M}}}_{\rm{in}}$ the mass accretion rate that crosses r$_{\rm{in}}$. \\

Substituting equations~\ref{eq2}-\ref{eq5} in \ref{eq1}, the viscous time scale is:
\begin{equation}
\rm{t}_{\rm{visc}} \sim 5\times10^{3} \ \alpha_{0.1}^{-4/5} \ {\dot{\rm{M}}}_{\rm{in},25}^{-3/10} \ {{\rm{M_{CO}}}_{,5}}^{1/4} \ {{\rm{r_d}}_{,10}}^{5/4} \ s,
\label{eq6}
\end{equation}
where $\alpha_{0.1} = \alpha / 0.1$, ${\dot{\rm{M}}_{\rm{in},25}}={\dot{\rm{M}}}_{\rm{in}} / ({10^{25} {\rm{g \ s^{-1}}}})$, ${{\rm{M_{CO}}}_{,5}}={{\rm{M_{CO}}}} / ({5 M_{\odot}})$, and ${{\rm{r_{d}}}_{,10}}={{\rm{r_{d}}}} / ({10^{10} {\rm{cm}}})$.

Since the CO is either a 5M$_{\odot}$ BH, or a 1.4M$_{\odot}$ NS, then the viscous time scale for a thin accretion disk (with a typical value of $\alpha$=0.1) will depend solely on r$_{\rm{d}}$ and ${\dot{\rm{M}}}_{\rm{in}}$. The mass accretion rates in our simulations were within (10$^{25}-$10$^{26}$)g~s$^{-1}$ and (10$^{24}-$10$^{25}$)g~s$^{-1}$ for the BH and NS models, respectively (see Figures~\ref{fig4} and \ref{fig5}). Thus, the viscous time for an accretion disk between $r_{\rm{d}}\sim$10$^9$-10$^{10}$~cm, is of order $\rm{t}_{\rm{lag}} \sim$10$^3$-10$^4$~s. 

Meanwhile, for the case when the material \textcolor{black}{that crossed the inner boundary free-falls directly onto the CO}, we have:
\begin{equation}
\rm{t}_{\rm{ff}} \approx \sqrt{\frac{ {\rm{r_{in}}}^3}{{\rm{G}} \ M_{CO}} } \sim 10^{3} \ {{\rm{r_{in}}}_{,11}}^{3/2} \ {{\rm{M_{CO}}}_{,5}}^{-1/2} \ s,
\label{eq7}
\end{equation}
where G is the gravitational constant, and ${{\rm{r_{in}}}_{,11}}={{\rm{r_{in}}}} / ({10^{11} {\rm{cm}}})$. 

The case when the material \textcolor{black}{that crosses r$_{\rm{in}}$ and} free-falls onto an accretion disk \textcolor{black}{(with r$_{\rm{d}}<$r$_{\rm{in}}$),} and then spirals-in through the accretion disk until it reaches the CO, is a combination of the results found in Equations~\ref{eq6}-\ref{eq7}. The lag time for a disk with a radius of r$_{\rm{d}}$=10$^9$cm disk (``small disk") will be t$_{\rm{lag}} \sim$10$^3$s, while it will be an order of magnitude larger for a disk with r$_{\rm{d}}$=10$^{10}$cm (``large disk"). Applying the correspondent t$_{{\rm lag}}$ in the ${\dot{\rm{M}}}_{\rm{in}}$ (for both the small and large disk), \textcolor{black}{produces a difference in ${\dot{\rm{M}}}_{\rm{in}}$ of at most 5\% and 45\% for the small and large disks (compared to the correspondent models with no lag time). We must note that the error for a large disk is within the variability produced by the self-regulated jet (see Figures~\ref{fig4} and \ref{fig5}).}

In order to verify that the error due to the omission of the lag-time for the case of a small disk is very small, we ran an extra set of models akin to models BH0.05 and NS0.05, but taking into account a lag time equal to $\rm{t}_{\rm{lag}}$=10$^3$s (models BH0.05lag and NS0.05lag, see Table 1 for further details). Comparing the ${\dot{\rm{M}}}_{\rm{in}}$ of models BH0.05lag and NS0.05lag with their respective case with t$_{\rm{lag}}$=0, we find that \textcolor{black}{the difference between the two mass accretion rates is at most $\pm$5\%}, confirming the previous error calculation for a small disk. Thus we can conclude that the error produced by omitting the lag time is in many cases inconsequential (for small disk), or is within the variability produced by the \textcolor{black}{self-regulated jet} (for large disks). Hence, the omission of the lag-time does not have \textcolor{black}{major} repercussions in the powering of the self-regulated jet in our models.

\bsp	
\label{lastpage}
\end{document}